\documentclass[pra,preprint,showpacs,superscriptaddress]{revtex4}
\usepackage{psfrag,amsmath,amssymb,amsfonts,bbm,latexsym,color,dcolumn}
\usepackage[dvips]{graphicx}
\begin{document}
\title{Systematic Correction for ``Demonstration of the Casimir Force in the 0.6 to 6 $\mu$m Range"}

\author{S.K. Lamoreaux}
\affiliation{Yale University, Department of Physics, P.O. Box 208120, New Haven, CT 06520, USA}

\date{\today}

\begin{abstract}
A new systematic correction for Casimir force measurements is proposed and applied to the results of an experiment that was performed more than a decade ago.  This correction brings the experimental results into good agreement with the Drude model of the metallic plates' permittivity.  The systematic is due to time-dependent fluctuations in the distance between the plates caused by mechanical vibrations or tilt, or position measurement uncertainty, and is similar to the correction for plate roughness.   
\end{abstract}

\pacs{12.20.Fv, 11.10.Wx, 73.40.Cg, 04.80. Cc}

\maketitle

\section{Introduction}

This is a short note reporting a new type of correction to the Casimir force \cite{Casimir}, and the results of its application to my experiment that was performed some 15 years ago \cite{Lamoreaux}. One might question whether it is worth reanalyzing an experiment that is so very dated, however, this work stands, together with our work with Germanium \cite{kim}, as the experiments with the largest plate separations, and are particularly sensitive to a number of fundamental, thermal, and systematic effects.  Indeed, \cite{Lamoreaux} led to a resurgence on interest in the Casimir measurement field \cite{Casexp}, and has been discussed in works of varying sophistication, from those presenting revolutionary new ideas \cite{Bostrom} to those that indicate that the authors cannot read a graph \cite{xxxx}. 

Although my experiment \cite{Lamoreaux} was intended as a demonstration, the deviation between its results and the theory presented by B\"ostrom and Sernelius \cite{Bostrom} remains as a puzzle.  Despite years of theoretical work and years of questioning, there has been no satisfactory explanation of the discrepancy. In particular, it would appear that the Drude model must be the correct one to describe the plates, because the so-called Plasma model implies a superconducting boundary condition at zero frequency.  

Unfortunately, the raw data that led to \cite{Lamoreaux} is no longer available, however there is enough information in that paper to make a good estimate of the correction.  As the correction depends on external factors that were not measured, it is not clear that the raw data would help much in any case.  The work    
reported here was inspired by our recent and ongoing re-measurement of the Casimir force between Au plates, which support the Drude model better than the Plasma model. In the course of our recent work, it occurred to us that time-dependent fluctuations between the plates can lead to a correction, much like the surface roughness correction that has been studied by Prof. Mostepanenko and collaborators \cite{xxxx}.  

The interesting features of \cite{Lamoreaux} are as follows:  The Drude model appears to better describe the long-distance (greater that 1.5 $\mu$m) data; The short-distance data appears to agree with the Plasma model prediction.   The roughness usually associated with optical surfaces is too small to account for the deviation between the Drude theory and experiment, and seems to have the wrong form as in this case, the discrepancy should falls, as a fraction of the force, as $1/d^2$, where $d$ is the separation between the plates.  Thus for the Casimir force alone, the effect should be very short range.  In \cite{Lamoreaux}, a background potential existed, creating an electrostatic force that was greater than the Casimir force over the measurement range.  Thus, the effects of both forces must be considered together.

\section{Corrections due to Vibration and Distance Calibration Uncertainties}

The force between two plates, for small variations $\delta(t)$ of the distance $d$ is 
\begin{equation}
F(d+\delta(t))=F(d)+F'(d)\delta(t) +{1\over 2} F''(d)\delta^2(t).
\end{equation}
If we assume that $\delta(t)$ represents a stationary random process with zero mean, there are two cases to consider.  First, if the correlation time of $\delta(t)$ implies frequencies higher than the measurement bandwidth, the term linear in $\delta(t)$ does not contribute to anything, while the second order term results in a change in the apparent force,
\begin{equation}\label{fcorr}
F_a(d)=F(d)+{1\over 2} F''(d)\langle \delta^2 \rangle
\end{equation}
where $\langle \delta^2(t)\rangle =\delta_{rms}^2$ is the mean-square fluctuation. It should be noted that $\delta_{rms}$ can have contributions from multiple sources, which, if uncorrelated, can be added in quadrature. In addition, a finite surface roughness can be included here; the form of Eq. (\ref{fcorr}) does not distinguish between spatial or temporal roughness.

Second, if the fluctuations frequency is within the measurement bandwidth, which is the case for uncertainties in the distance determination, there will be an excess scatter associated with the force measurement, 
\begin{equation}\label{sigcorr}
\sigma_{F_a}^2=\sigma_F^2+ (F'(d)\delta_{rms})^2
\end{equation}
while the apparent average force is given by Eq. (\ref{fcorr}), as before.  

\section{Application to the Experiment}

In \cite{Lamoreaux}, there is a large background electrostatic force that is used for determining the absolute distance, obtained by fitting to $\beta/(d-d_0)$, using points at distances greater than 2 $\mu$m.  For the approximations here, let us assume that there are no significant corrections at these long distances.  Based on Fig. (3) of \cite{Lamoreaux}, the background electrostatic force is
\begin{equation}
F_e(d)={\beta\over d};\ \ \ \ \beta=(215\pm 7)\ {\rm \mu dyne\ \mu m}.
\end{equation}
The total force is the electric plus Casimir force,
\begin{equation}
F(d)=F_e(d)+F_c(d)
\end{equation}
and the first derivative at $d=0.62\ \mu$m is approximately $1000\ \mu$dyne$/\mu$m.  Comparing the size of the error bars in Fig. 4 of \cite{Lamoreaux}, where the 0.1 $\mu$m bins have $1/10$ the data of the 1 $\mu$m bins and should be $\sqrt{10}$ time larger.  It is observed that they are $1.32\sqrt{10}$ larger, so the contribution to the error due to fluctuations is 4 $\mu$dyne.  This implies that the rms fluctuations in position on a time scale of the measurement of each point (50 seconds) is 40 nm.  This is to be compared to 14 nm in our present experiment, where the intrinsic signal to noise is similar, as is the applied calibration voltage.  The excess noise in \cite{Lamoreaux} is likely due to the faster rate of drift in position.  It is stated in \cite{Lamoreaux} that the fluctuations in absolute position measurement is less than 100 nm, which is consistent with the result here. On the other hand, the quality of the data in Fig. (3) of \cite{Lamoreaux} suggest that perhaps 40 nm is optimistic; unfortunately the original data set is not available to further investigate this point.  For fluctuations at this level, the apparent change in force is less than 5\% so these fluctuations do not contribute to the discrepancy.

Our recent work shows that, due to vibrations and tilt, there is an rms position fluctuation of 20 nm in a .01 to 5 Hz bandwidth for our torsion pendulum supported by a tungsten wire of a few cm length. The variations at low frequency are dominated by tilt of the pendulum, and correspond to angular fluctuations of order $1\times 10^{-7}$ radians.  Given that the pendulum in the present experiment has a length of 4 cm compared to an effective length of nearly 80 cm for \cite{Lamoreaux}, we might expect position rms fluctuations of order 400 nm, as the change in position is roughly the pendulum length time the tilt angle, which is not to be confused with the torsional motion angle.  However, the bandwidth of the swinging mode of the longer pendulum is lower (it is relatively smaller by a approximately a factor $1/\sqrt{40}=1/6.3$, so the effective noise should be a factor $1/\sqrt{6.3}=1/2.5$) implying an rms noise of 400 nm$/2.5=160$ nm.  We can therefore take $\delta_{rms}=100$ nm as a lower limit.  In operating the experiment \cite{Lamoreaux}, the problems with tilt noise were about an order of magnitude worse than our present experiment, consistent with relative size of the rms position fluctuations as discussed here.  

Only angular fluctuations at frequencies below the swinging mode frequency will contribute significantly to the relative separation fluctuation between the plates because the magnetic damper tends to stabilize the relative position of the plates.  On the other hand, vibrations that cause a net translational motion of the system couple in a different way, and frequencies above the mode frequency can contribute.  The angular noise dominates so we neglect translational vibrations in this discussion.

Let us now consider the combined effect of $F_e(d)$ and $F_c(d)$ for either the Plasma or Drude models of permittivity. The correction to the force is given by Eq. (2), as
\begin{equation}
F_a(d)-F(d)={1\over 2}(F_e''(d)+F''_c(d))\delta_{rms}^2.
\end{equation}  
It's easy to calculate $F_e''(d)$, while $F''_c(d)$ can be numerically evaluated.  The Plasma model and Drude model forces were calculated using tabulated Au properties, with the interesting result that the second derivative of the Drude model is about twice as large, in the 0.5 to 1 $\mu$m range, as that for the Plasma model, which isn't too surprising as the Drude model force is falling off more rapidly  in this region.  It should also be noted that $F''_e\propto 1/d^3$ so there is a constant offset for large $d$, when the correction is multiplied by $d^3$ as shown in the graph.  

The results are shown in Fig. 1, where it is clear that the Drude model has much better agreement. Furthermore, both the large and small distance data agree with the theory, unlike the case of any other model. As an aside, a re-measurement of the radius of curvature of the spherical plate used in \cite{Lamoreaux} shows $R=12.4\pm0.1$ cm, which has lower error than the number reported in the Erratum \cite{Lamoreaux}.

The agreement can be made better by allowing $\delta{rms}$ vary with distance.  It is reasonable to assume that the runs which attained the lowest separation were obtained when the system and environment was particularly quiet.  For example, if we let $\delta_{rms}=\sqrt{{d\over 3 {\rm\mu m}}}\ \mu$m then $\chi^2=0.79$ for the Drude model, while $\chi^2=6.9$ for the Plasma model.

\begin{figure}[t]
\includegraphics[width=1.0\columnwidth,clip]{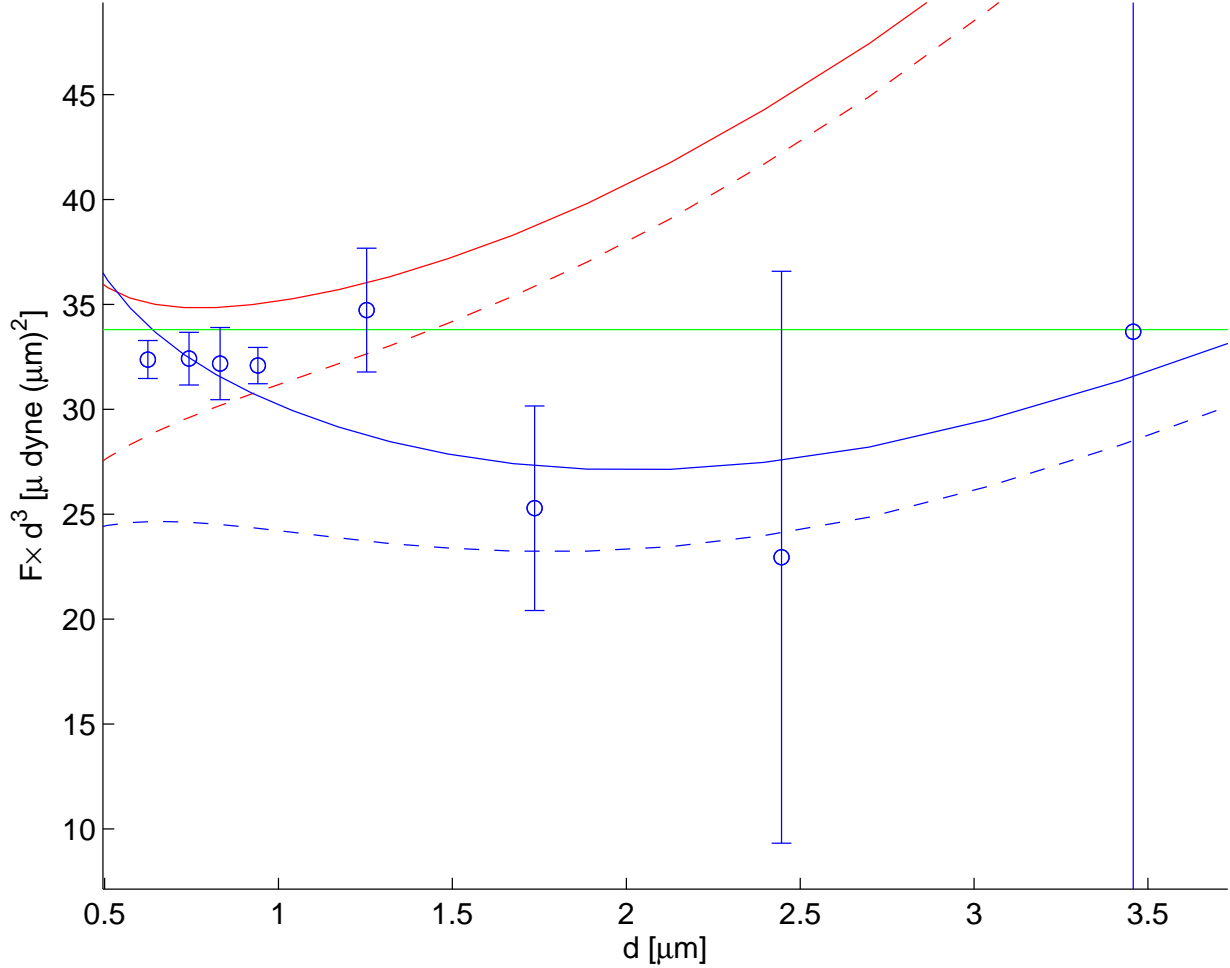}
\caption{Results taking $\delta_{rms}=100$ nm. Green line: perfect conductor, zero temperature;  solid red: Plasma model with distance fluctuations; dashed red: Plasma model without distance fluctuation correction; solid blue: Drude model with distance fluctuations; dashed blue: Drude model without distance fluctuation correction.  Reduced $\chi^2=6.17$ (prob. $< 10^{-5}$, 6 d.o.f.) for the Plasma model with distance correction, while $\chi^2=1.75$ for the Drude model with distance correction (prob. 10\%, 6 d.o.f.). }
\label{fig1}
\end{figure}

\section{Discussion and Conclusion}

By assuming that the relative separation between the plates of a Casimir experiment is fluctuating on a short time scale, the results given in \cite{Lamoreaux} can be brought into agreement with theory.  The required rms fluctuations appear as large, of order 100 nm, but such a level is not unreasonable.  Given that the total pendulum length is nearly one meter, a tilt of $10^{-7}$ radians is all that is required to generate the required fluctuations.  Such level of tilt is easily generated by air currents moving past the apparatus, and by unavoidable oscillations of the floor.  It should be noted that the rms position fluctuations are due mainly to angle fluctuations with frequency less than about 0.5 Hz, the natural frequency of the pendulum's swinging mode.  Measurements with our new apparatus at Yale shows 20 nm rms fluctuations, however, the pendulum is only a few cm in length.   A full analysis of the pendulum and how various tilts and vibrations affect the relative positions of the plates is a tedious but elementary exercise.  We do note however that within the feedback bandwidth (which is greater than the measurement bandwidth), the system compensates for a tilt by adjusting the torsion pendulum angle, in order to keep the differential capacitor balanced.  Thus, for frequencies below about 0.1 Hz, a tilt position offset is approximately doubled for the interplate separation.   Indeed, the extreme sensitivity of the apparatus to tilt and vibration was readily apparent; in order to take sensible data, the experiment could only be operated between 11 pm and about 5 am, and the air conditioning ducts into the room had to be blocked. The required rms position fluctuation of 100 nm is below the minimum plate separation of 600 nm, and falls into what can be considered a reasonable range.  The angular fluctuations appeared as a very slow drift, causing a changing in the distance offset between the plate, and was of order 1 $\mu$m/hour.  On top of this slow drift, according to the calculations here, were rapid fluctuations with periods up to 2 second (the frequency of the pendulum swinging mode) with rms deviations of order $10^{-7}$ rad in the .01-.5 Hz bandwidth.  

It appears that $\delta_{rms}=100$ nm gives close to a best fit, and thus should be considered as an adjustable parameter, the value of which is verified by other means described in this note.  Also to be noted that other contributions to $\delta_{rms}$ can be included by adding all contributions in quadrature.  Of course, other systematic effects can contribute, such as a contact potential that varies with distance.  The contact potential was not measured as a function of distance in \cite{Lamoreaux}, however, our recent work with Au suggests that the contact potential is nearly constant.  This does not preclude the possibility that there was such a variation in \cite{Lamoreaux}.  

If $\delta_{rms}(d)$ depends on distance, which is a very distinct possibility, it is easy to see that the the data can be brought into better agreement with the Drude model in particular.  Certainly, $\delta_{rms}$ depends on time, and the data runs that attained the closest separation were likely obtained when the system and the environment was the quietest.  This level of fine tuning is beyond the scope of the brief analysis presented here, and beyond the scope of credibility.  

Effects of vibration are important for all Casimir experiments.  Even in the absence of external perturbations, AFM cantilevers, for example, exhibit Brownian motion and the effects of such need to be taken into account.  The implication is that very stiff springs should be used.  In the case of the torsion pendulum experiment,  feedback is used to keep the torsion angle fixed; this reduces position fluctuations   due to Brownian motion, but can make the system more sensitive to vibrations and tilts, as discussed above.

The apparent force due to the large $1/d$ electrostatic force varies as $1/d^3$, with magnitude relative to the Casimir force (perfect conductors) of $\beta \delta_{rms}^2/2= 1.2 \ \mu$dyne $\mu$m$^3$, about 3\% of the perfect conducting force.  

Again, the work described in \cite{Lamoreaux} was presented as a demonstration; the analysis here shows that there is a possible systematic effect that can lead to a substantial increase in the apparent Casimir force.  In this case, the large electrostatic force, present for calibration, contributes substantially, particularly at large separations.  Because its contribution scales as $1/d^3$, it appears as a scale factor for the Casimir force, for distances around 0.5 $\mu$m.  Our recent work at Yale led to the consideration of these effects, and also appears to support the Drude model for the permittivity.  We hope to complete these studies in the very near future.  

Finally, it must be emphasized that \cite{Lamoreaux} was intended as a demonstration; the results presented here should not be considered as a verification of the Bostr\"om/Sernelius theory \cite{Bostrom}, or as evidence against the Plasma model, but the discovery of a systematic effect that brings the experimental results into agreement with the theory described in \cite{Bostrom}. It is unclear whether additional systematic effects exist, however, it had always been my impression that my experimental result was likely contaminated by additional, possible large systematics \cite{patch}.  I have never considered the results of this experiment as suitable for constraining possible new long range forces; had I felt such was meaningful I would have produced those limits in the context of \cite{Lamoreaux}.  Here I have presented what I consider a very likely systematic effect.  We now know to pay careful attention to position fluctuations in our ongoing work.  

\section{Acknowledgements}

The author thanks Alex Sushkov for helpful discussions.

This was funded by Yale University, and DARPA/MTO's Casimir Effect Enhancement
project under SPAWAR contract no. N66001-09-1-2071.

\end{document}